\documentclass[twocolumn]{emulateapj}
\shorttitle{Physical Contact  between the +20 km s$^{-1}$ Cloud and the CND}
\shortauthors{Takekawa et al.}
\begin{document}

\title{Physical Contact  between the +20 km s$^{-1}$ Cloud and the Galactic Circumnuclear Disk}

\author{Shunya Takekawa$^1$, Tomoharu Oka$^{1,2}$, and Kunihiko Tanaka$^2$}
\affil{
$^1$School of Fundamental Science and Technology, Graduate School of Science and Technology, Keio University, 3-14-1 Hiyoshi, Yokohama, Kanagawa 223-8522, Japan\\
$^2$Department of Physics, Institute of Science and Technology, Keio University, 3-14-1 Hiyoshi, Yokohama, Kanagawa 223-8522, Japan
}
\email{shunya@aysheaia.phys.keio.ac.jp}

\begin{abstract}
This paper reports the discovery of evidence for physical contact between the Galactic circumnuclear disk (CND) and { an exterior} giant molecular cloud.
The central 10 pc of our Galaxy has been imaged in the HCN {\it J}=1--0, HCO$^+$ {\it J}=1--0, CS {\it J}=2--1, H$^{13}$CN {\it J}=1--0, SiO {\it J}=2--1, SO {\it N$_J$}=2$_3$--1$_2$, and HC$_3$N {\it J}=11--10 lines using the Nobeyama Radio Observatory 45 m radio telescope.
Based on our examination of the position--velocity maps of several high-density probe lines, we have found that an emission ``bridge" may be connecting the +20 km s$^{-1}$ cloud (M--0.13--0.08) and the negative-longitude extension of the CND.
Analyses of line intensity ratios imply that the chemical property of the bridge is located between the +20 km s$^{-1}$ cloud and the CND.
{ We introduce a new interpretation that a part of the CND may be colliding with the 20 km s$^{-1}$ cloud and the collision may be responsible for the formation of the bridge.}
Such collisional events could promote mass accretion onto the CND or into the inner ionized cavity, { which may be further tested by proper motion studies.}
\end{abstract}


\section{Introduction}
Most massive galaxies are believed to harbor supermassive black holes (SMBHs), some of which are observed as enormously bright compact objects, that is, active galactic nuclei. 
Our Galaxy is thought to harbor a $4\times10^6$ $M_{\odot}$ SMBH at its dynamical center (Ghez et al. 2008; Gillessen et al. 2009), which is recognized as the compact nonthermal radio source Sgr A$^*$ (Balick \&\ Brown 1974).
Although Sgr A$^*$ is an extremely dim nucleus despite its huge mass, several X-ray studies have suggested its past activities (e.g., Koyama et al. 1996, 2008).

Sgr A$^*$ lies in the center of the ionized gas ``minispiral" (Lo \& Claussen 1983; Zhao et al. 2009, 2010), which is encompassed by a dense molecular gas ring, i.e., the ``circumnuclear disk" (CND).
The CND has an inner radius of $\sim 2$ pc, is inclined $\sim 20\arcdeg$ to the Galactic plane, and has a rotational velocity of $\sim 110$ km s$^{-1}$ (Genzel et al. 1985; Jackson et al. 1993).  
{ The CND extends up to $\sim 7$ pc from the center toward negative Galactic longitude and up to $\sim 3$ pc toward positive longitude (Serabyn \& G\"{u}sten. 1986; Sutton et al. 1990).
 It consists} of the 2 pc ring (e.g., Christopher et al. 2005; Montero-Casta\~{n}o et al. 2009; Mart\'{i}n et al. 2012) and the negative-longitude extension (NLE; Oka et al. 2011).

The CND is considered to be the mass reservoir for the central activities.
The age of the CND is suggested to be younger than $\sim$10$^6$ years (e.g., Requena-Torres et al. 2012).
The CND may have been formed by the tidal capture and disruption of a giant molecular cloud (GMC) by the central SMBH within 10$^6$ years (Sanders 1998; Wardle \& Yusef-Zadeh 2008; Liu et al. 2012; Mapelli et al. 2012; Mapelli \& Trani 2016).
Indeed, two GMCs, M--0.13--0.08 and M--0.02--0.07 (+20 and +50 km s$^{-1}$ clouds, respectively), are located on the southwest and southeast sides of the CND in Galactic coordinates, respectively.
These clouds are distributed in the projected distance range of $\sim3$--15 pc from Sgr A$^*$.
The +20 km s$^{-1}$ cloud presumably lies in the foreground of Sgr A$^*$, because it appears as a dark patch against the 2 $\mu$m radiation of the central stellar cluster (e.g., G\"{u}sten \& Henkel 1983).

Several authors have suggested the possibility of interaction between the +20 km s$^{-1}$ cloud and the CND.
Okumura et al. (1989, 1991) reported on a ``finger-like extension" from the +20 km s$^{-1}$ cloud toward the CND.
This long filamentary structure, which is also referred to as a ``southern streamer", was suggested to be feeding the CND (Ho et al. 1991, Coil \& Ho 1999, 2000; Lee et al. 2008). 
However, it remains unclear whether the streamer actually connects the +20 km s$^{-1}$ cloud to the CND (Herrnstein \& Ho 2005).
Minh et al. (2013) reported a widespread negative-velocity component at the location of the +20 km s$^{-1}$ cloud.
They suggested that this component may feed the CND but is probably not a part of the +20 km s$^{-1}$ cloud.
Thus, no direct connection between the +20 km s$^{-1}$ cloud and the CND has been detected yet.

On the other hand, a recent gas kinematic model has suggested that the +20 and +50 km s$^{-1}$ clouds may be located $\sim 60$ pc apart from Sgr A$^*$ (e.g., Kruijssen et al. 2015).
It has been unclear whether the GMCs are interacting with the CND or not.
{ Examination of the detailed gas kinematics and molecular abundance based on appropriate spatial resolution observations may provide new aspects of the physical relation between the CND and the ambient molecular clouds.}
In this paper, we report the discovery of an emission feature that bridges the +20 km s$^{-1}$ cloud and the CND in latitude--velocity maps of millimeter-wave molecular lines.
{ We describe the new observations in Section 2 and present the results in Section 3.
In Section 4, we discuss the property of the emission bridge, introducing the new scenario that the asymmetric part of the CND may be colliding with the 20 km s$^{-1}$ cloud.
We summarize this study briefly in Section 5.}

\section{Observations}
On-the-fly (OTF) mapping observations were performed with the Nobeyama Radio Observatory (NRO) 45 m radio telescope during 2014 February 5--12 and March 28--30.
The mapping area was $6\arcmin \times 6\arcmin$ ($-0.11\arcdeg\!\leq\!l\!\leq\!-0.01\arcdeg$ and $-0.11\arcdeg\!\leq\!b\!\leq\!-0.01\arcdeg$), covering the entire CND.
Target lines were HCN {\it J}=1--0, H$^{13}$CN {\it J}=1--0, HCO$^+$ {\it J}=1--0, CS {\it J}=2--1, SiO {\it J}=2--1, SO {\it N$_J$}=2$_3$--1$_2$, and HC$_3$N {\it J}=11--10, which were selected based on our 3 mm band line surveys (Takekawa et al. 2014).
According to our classification, the HCN, H$^{13}$CN, HCO$^+$, and SiO lines belong to the CND type, the CS and SO lines belong to the hybrid type, and the HC$_3$N line belong to the GMC type.
The CND and GMC types are the lines that mainly trace the CND and the GMCs, respectively.
The hybrid type possesses characteristics of both the CND and GMC types.

The TZ1 V/H receivers were operated  in the two-sideband mode.
The half-power beamwidth (HPBW) and the main-beam efficiency ($\eta_{\rm MB}$) at 86 GHz were $20\arcsec$ and 0.4, respectively.
The SAM45 spectrometer was operated in the 1 GHz bandwidth (244.14 kHz resolution) mode.  
The system noise temperature ($T_{\rm sys}$) ranged from 150 to 330 K during the observations.
The emission-free reference positions at $(l, b)=(0.0\arcdeg, +0.5\arcdeg)$, $(0.0\arcdeg, -0.5\arcdeg)$ were observed alternately.
Pointing errors were corrected every 1.5 hr by observing the SiO maser source VX Sgr at 43 GHz with the H40 receiver.  
The pointing accuracy for both azimuth and elevation was better than 3\arcsec\ (rms).
The standard chopper-wheel method was used to calibrate the antenna temperature.

All data were reduced using the NOSTAR reduction package developed at the NRO.
We used linear fittings to subtract the baselines of all the obtained spectra.
The maps were convolved using Bessel--Gaussian functions and resampled onto $7.5\arcsec\times7.5\arcsec\times2$ km s$^{-1}$ regular grids.  The data in the antenna temperature ($T_{\rm A}^*$) scale were converted to the main-beam temperature ($T_{\rm MB}$) scale by multiplying by $1/\eta_{\rm MB}$.  The rms noise level of the resultant maps was 0.1 K in $T_{\rm MB}$.

\section{Results}

\begin{figure}[!tbh]
\begin{center}
\includegraphics[width=70mm]{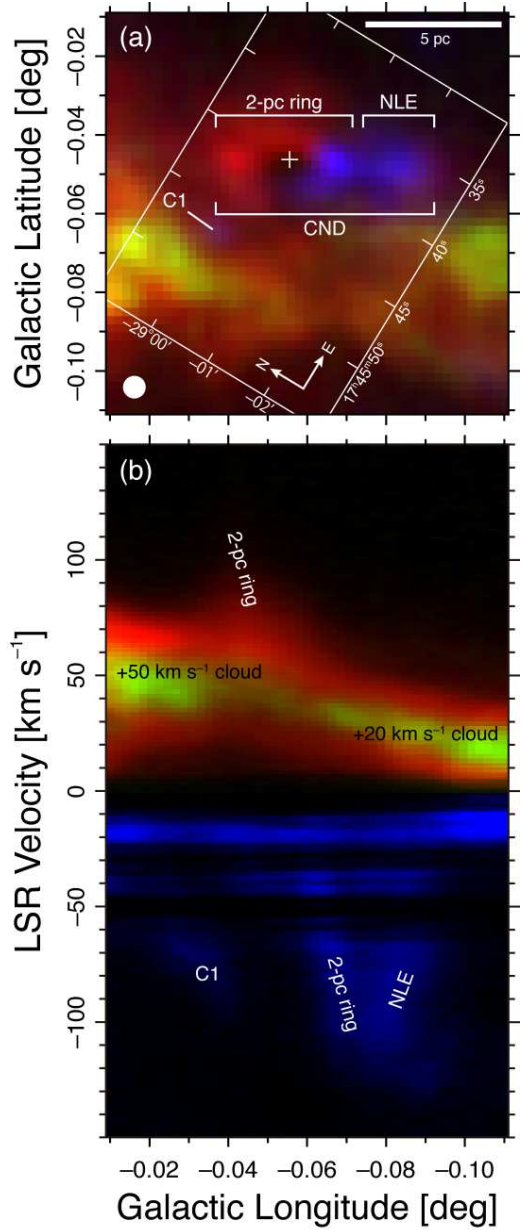}
\caption{(a) Composite velocity-integrated intensity map and (b) longitude--velocity ({\it l--V}) map integrated over latitude.
Red and blue indicate the redshift ($V_{\rm LSR} > 0$) and blueshift ($V_{\rm LSR} < 0$) in the HCN $J$=1--0 emission, respectively, and green indicates the HC$_3$N $J$=11--10 emission.  
The equatorial coordinate (J2000) is overlaid for comparison to relevant papers.
The ``+'' indicates the location of Sgr A$^*$. The white circle indicates the HPBW at 86 GHz ($20\arcsec$).
The integration velocity ranges from $-150$ to +150 km s$^{-1}$, and the integration latitude ranges from $-0.11\arcdeg$ to $-0.01\arcdeg$.  }
\end{center}
\end{figure}

Figure 1(a) is a composite image of the HCN {\it J}=1--0 and HC$_3$N {\it J}=11--10 lines.
The spatial distribution of the CND (the 2 pc ring and the NLE) is delineated as highly red- and blueshifted emissions.
A small hole at the Sgr A$^*$ position is attributed to absorption by the foreground gas against the intense continuum radiation.  
{ In the latitude-integrated longitude--velocity ({\it l--V}) map [Figure 1(b)], the positive-velocity part of the CND is almost buried in the two GMCs.
It protrudes} at $l\sim -0.04\arcdeg$ toward positive high velocities.
The negative-velocity part of the CND, including the NLE, is visible where $V_{\rm LSR}\lesssim -30$ km s$^{-1}$.
The steep velocity gradient of the 2 pc ring is due to its rapid rotation.
However, that of the NLE has the opposite sign, the origin of which is currently unresolved.
Straight horizontal lines at $V_{\rm LSR}\simeq 0,\,-30$, and $-50$ km s$^{-1}$ on the {\it l--V} map are absorption features of the foreground spiral arms in the Galactic disk.
The small cloud with a moderate velocity width and a steep velocity gradient at $(l, b, V_{\rm LSR})\simeq (-0.04\arcdeg , -0.06\arcdeg ,-70\rm~km~s^{-1})$ is the ``C1 cloud" (Oka et al. 2011) or Cloud A (Amo-Baladr\'on et al. 2011).

\begin{figure*}[!tbh]
\begin{center}
\includegraphics[scale=2.]{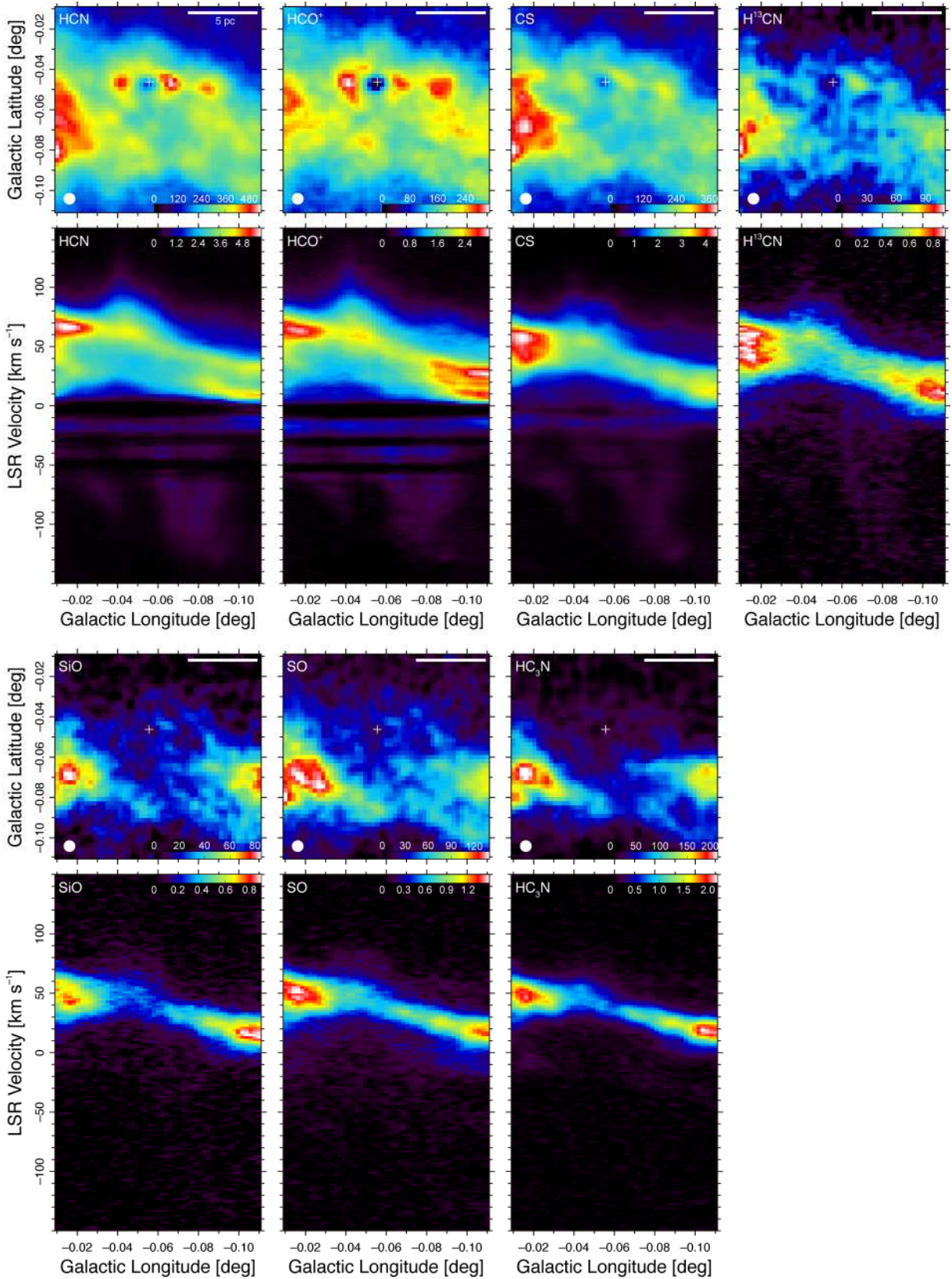}
\caption{
Velocity-integrated intensity maps and latitude-averaged {\it l--V} maps for all the observed lines (HCN {\it J}=1--0, HCO$^+$ {\it J}=1--0, CS {\it J}=2--1, H$^{13}$CN {\it J}=1--0, SiO {\it J}=2--1, SO {\it N$_J$}=2$_3$--1$_2$, and HC$_3$N {\it J}=11--10).
The ``+'' indicates the location of Sgr A$^*$.
The scale bar on the upper right corner indicates 5 pc at the Galactic center.
The white circle on the lower left corner indicates the HPBW at 86 GHz ($20\arcsec$).
The integration velocity ranges from $-150$ to +150 km s$^{-1}$, and the integration latitude ranges from $-0.11\arcdeg$ to $-0.01\arcdeg$.
The intensity units of the velocity-integrated maps and  {\it l--V} maps are K km s$^{-1}$ and K in the $T_{\rm MB}$ scale, respectively.
}
\end{center}
\end{figure*}

\begin{figure*}[!tbh]
\begin{center}
\includegraphics[scale=1.9]{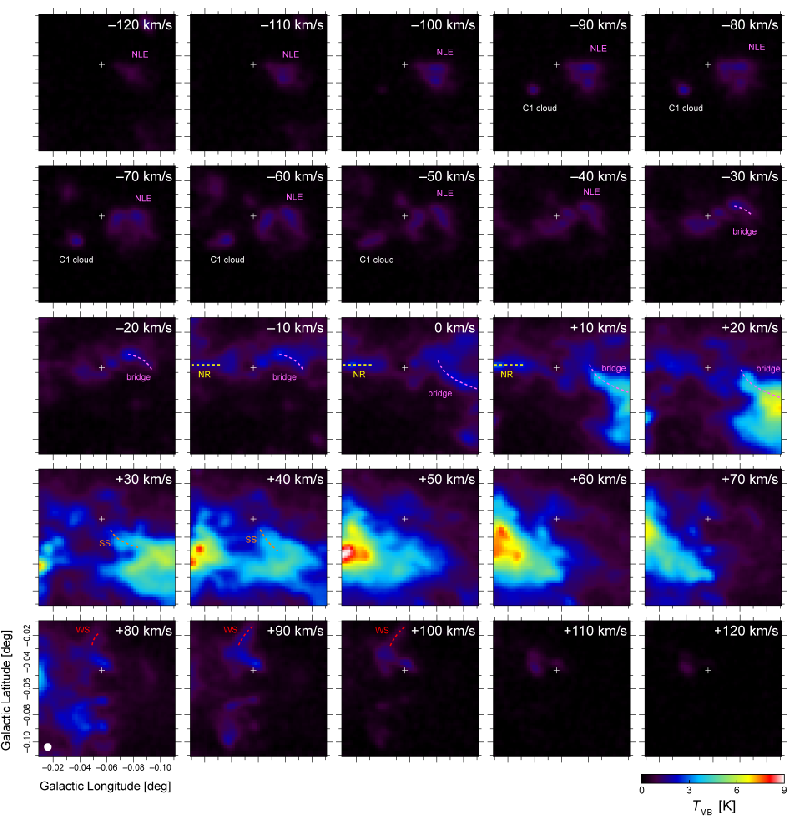}
\caption{
Velocity channel maps of the CS {\it J}=2--1 line emission.
The panels are arranged every five channels (10 km s$^{-1}$ intervals).
The peak intensity is 9.2 K in the $T_{\rm MB}$ scale.
The ``+'' indicates the location of Sgr A$^*$.
The white circle on the lower left corner indicates the HPBW at 86 GHz ($20\arcsec$). 
The magenta dashed lines trace the ``bridge".
The orange, red, and yellow dashed lines indicate the emission from the previously known streamers: the Southern Streamer (SS), the Western Streamer (WS), and the Northern Ridge (NR), respectively.
}
\end{center}
\end{figure*}

Figure 2 shows the velocity-integrated intensity maps and the latitude-averaged longitude--velocity ({\it l--V}) maps of all the observed lines.
The HCO$^+$ and HCN lines suffer from strong absorption while the other observed lines do not seriously.
{ We can see two emission layers of the GMCs at $(l, V_{\rm LSR})\simeq (-0.1\arcdeg , +10\rm~km~s^{-1})$ and $(-0.1\arcdeg , +30\rm~km~s^{-1})$ in the {\it l--V} maps of the HCN and HCO$^+$ lines.}
These structures are presumably derived from self-absorption in highly optically thick gas, which can be well traced by the HC$_3$N line (see also Figure 1).
The CND is visible in the maps of the HCN, HCO$^+$, CS, and H$^{13}$CN lines, and less (or partially) visible in the SiO and SO lines.

Figure 3 shows the velocity channel maps of the CS {\it J}=2--1 line.
The intense CS line is detected from both of the CND and the GMCs, and it is less affected by foreground absorption than the HCN and HCO$^+$ lines (see also Figure 2).
Thus, the the CS line is superior for probing the kinematics of molecular gas in the circumnuclear region.
We can see the previously known gas streamers, such as the Southern Streamer, the Western Streamer, and the Northern Ridge (Okumura et al. 1989, 1991; McGary et al. 2001; Herrnstein \& Ho 2002, 2003, 2005; Liu et al. 2012, 2013; see also \S 4.1).

\begin{figure*}[!tbh]
\begin{center}
\includegraphics[scale=1.6]{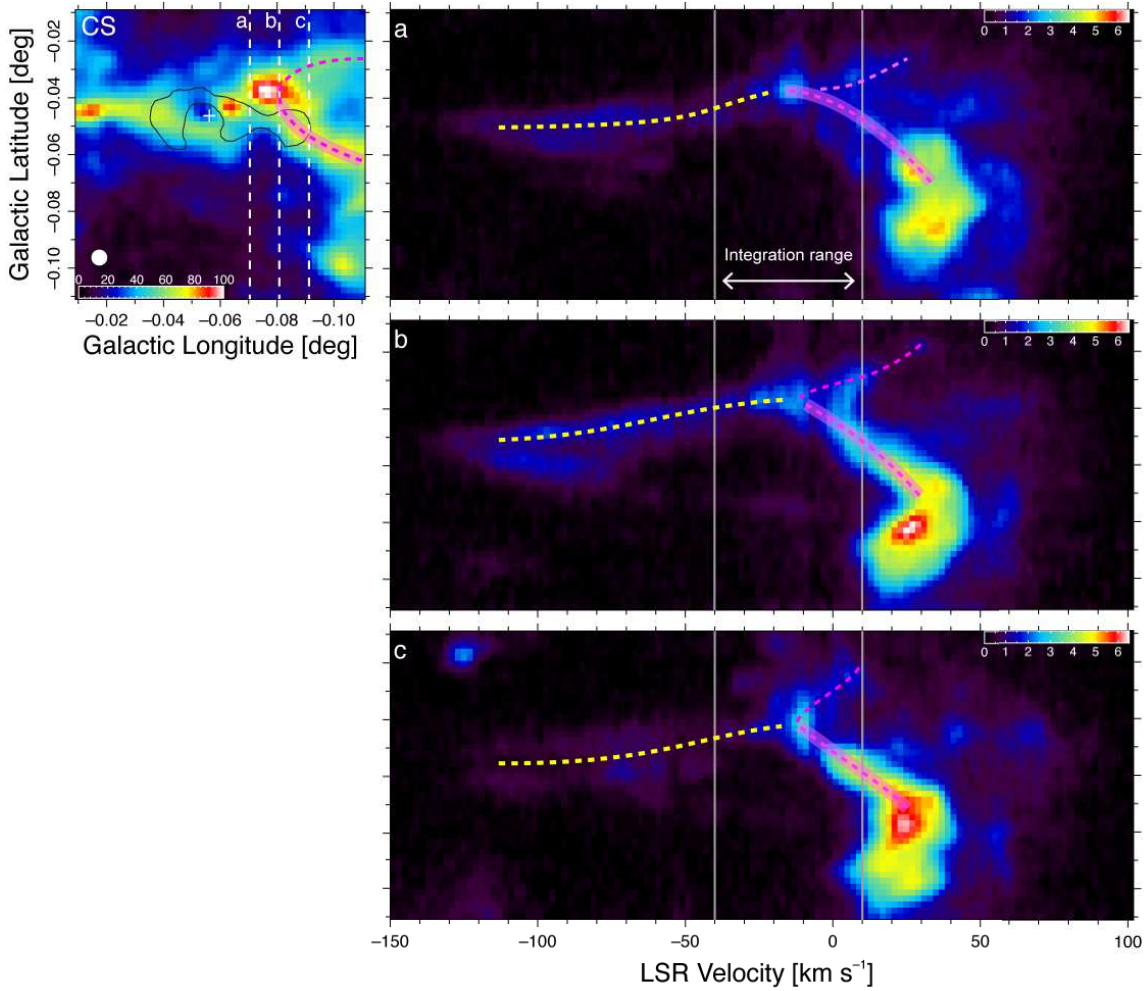}
\caption{
CS $J$=2--1 map integrated over the LSR velocity from $-40$ to $+10$ km s$^{-1}$ (upper left panel) and latitude-velocity ({\it b--V}) maps along three dashed white lines (a, b, c).
The ``+'' indicates the location of Sgr A$^*$.
The white circle on the lower left corner indicates the HPBW at 86 GHz ($20\arcsec$).
The black contour shows where the HCN {\it J}=1--0 velocity-integrated intensity is 350 K km s$^{-1}$, drawn to indicate the distribution of the CND.
The solid thick magenta lines and the dashed magenta lines trace the ``bridge" and the ``arc", respectively. 
The dashed yellow lines indicate the emission from the CND.
The intensity units of the velocity-integrated map and the {\it b--V} maps are K km s$^{-1}$ and K in the $T_{\rm MB}$ scale, respectively.
}
\end{center}
\end{figure*}

\begin{figure*}[!tbh]
\begin{center}
\includegraphics[scale=2]{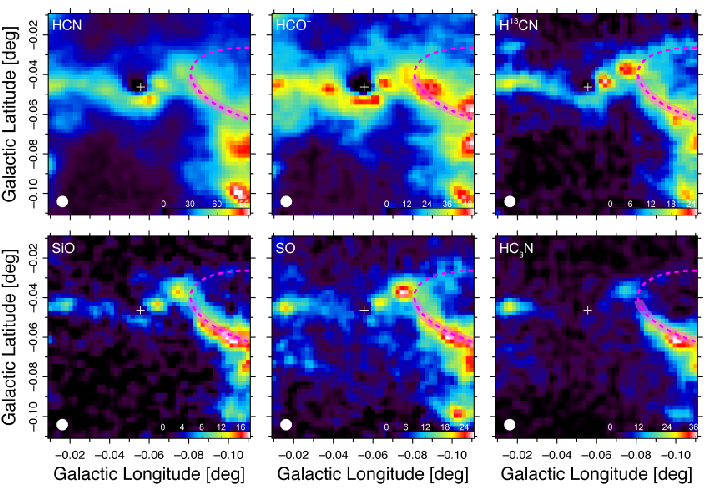}
\caption{
Intensity maps integrated over the LSR velocities from $-40$ to $+10$ km s$^{-1}$ in the HCN {\it J}=1--0, HCO$^+$ {\it J}=1--0, H$^{13}$CN {\it J}=1--0, SiO {\it J}=2--1, SO {\it N$_J$}=2$_3$--1$_2$, and HC$_3$N {\it J}=11--10 lines.
The ``+'' indicates the location of Sgr A$^*$.
The white circle on the lower left corner indicates the HPBW at 86 GHz ($20\arcsec$).
The solid thick magenta lines and the dashed magenta lines trace the ``bridge" and the ``arc", respectively. 
The intensity units are K km s$^{-1}$.
}
\end{center}
\end{figure*}

In these data, we noticed a ``bridge" of emission that connects the +20 km s$^{-1}$ cloud with the NLE in {\it l--b--V} space (Figure 3).
Figure 4 shows a map of the CS {\it J}=2--1 line emission integrated over the LSR velocities from $-40$ to $+10$ km s$^{-1}$ and three latitude-velocity ({\it b--V}) maps along three latitudinal slices.
The bridge (solid magenta lines in the figure) extends from $l\sim -0.11\arcdeg$ to $-0.08\arcdeg$.
In the {\it b--V} maps, the bridge originates from the +20 km s$^{-1}$ cloud at $(b, V_{\rm LSR})\simeq (-0.07\arcdeg, +30\, \mbox{km s}^{-1})$, moves upward with decreasing velocity, and then joins the NLE at $(b, V_{\rm LSR})\simeq (-0.04\arcdeg, -20\, \mbox{km s}^{-1})$.
There is a clump of intense emission at the point of contact, where an abrupt change in the velocity gradient takes place.
The longitudinal extent of this clump is similar to that of the NLE and traces its northern edge (see also Figure 3).
The projected length of the bridge is about 6 pc.  
The NLE in the {\it b--V} map is V-shaped, one line of which connects to the clump, and the other connects directly to the +20 km s$^{-1}$ cloud [Figure 4(b)].

The bridge and the clump at the contact point, together with the less intense emission at $b\sim -0.025\arcdeg$, form an arc-shaped feature with a radius of $\sim3$ pc (dashed magenta lines in Figure 4).
The bridge defines the lower-latitude rim of the arc, and the clump defines the eastern rim of the arc.
All these emission features also appear in the H$^{13}$CN, SiO, SO, and HC$_3$N maps but are less prominent in the HCN and HCO$^+$ maps (Figure 5), presumably because of severe foreground absorption (see also Figure 2).
Because these lines have high critical densities, the bridge, clump, and arc probably consist of dense [$n({\rm H}_2)\gtrsim 10^5$ cm$^{-3}$] molecular gas.

\section{Discussion}
\subsection{Previously reported features}
Previous studies have detected several features connected with the CND and their candidates.
In the following, we discuss those previously reported features that are also detected with our observations.

\begin{description}
 \item[Northern Ridge]
{A filamentary structure is located on the northeast side of Sgr A East (e.g., McGary et al. 2001).
This filament is clearly seen in the velocity channel maps at $-20$ km s$^{-1} \leq V_{\rm LSR} \leq +20$ km s$^{-1}$ (Figure 3), as well as the integrated intensity maps (Figure 5).
The HC$_3$N emission from the filament disappears at $l\simeq-0.04\arcdeg$, while that of the other lines seems to connect with the CND.
Because HC$_3$N is a UV-fragile molecule (Rodr\'{i}guez-Franco et al. 1998), the absence of the HC$_3$N may be understood as a result of the photodissociation by the intense UV field from the central cluster.
This suggests the proximity of the Northern Ride to the nucleus.}
\item[Western Streamer]
A long filament curves from the west to the north side of the 2 pc ring with a steep velocity gradient of 25 km s$^{-1}$ pc$^{-1}$ (McGary et al. 2001).
Submillimeter interferometric observations resolved the streamer into four arms (W-1 to W-4; Liu et al. 2012).
These arms could be tidally stretched and be connected to the 2 pc ring or penetrate inside the ring.
W-1 is seen in our HCN, H$^{13}$CN, HCO$^{+}$, SO, and CS maps.
It appears at +80 km s$^{-1} \leq V_{\rm LSR}\leq+100$ km s$^{-1}$, showing no velocity gradient in our maps (Figure 3).
This suggests that W-1 moves in parallel with the rotation of the CND with high angular momentum, as previously inferred by Liu et al. (2012).
Counterparts of W-2 and W-3 may be a diffuse component located at $b\sim-0.03\arcdeg$ in the velocity range of +20 km s$^{-1} \lesssim V_{\rm LSR}\lesssim+70$ km s$^{-1}$ (Figure 3).
W-4, which is located at $(l, b)\simeq(-0.08\arcdeg,-0.04\arcdeg)$, is likely to be a part of the NLE or the northern edge of the bridge.
\item[Southern Streamer]
A long filament is elongated from the +20 km s$^{-1}$ cloud to the southern edge of the 2 pc ring (e.g., Coil \& Ho 1999, 2000).
This appears at $b\sim-0.06\arcdeg$ with a velocity range of +20 km s$^{-1} \lesssim V_{\rm LSR}\lesssim+40$ km s$^{-1}$, showing a slight velocity gradient (Figure 3).
Note that the streamer stems from the same position as the root of the bridge at $(l, b, V_{\rm LSR})\simeq (-0.08\arcdeg , -0.07\arcdeg, +20\rm~km~s^{-1})$.
The streamer was suggested to be feeding the CND (Okumura et al. 1991; Ho et al. 1991; McGary et al. 2001).
However, it is still unclear whether the streamer is directly connected to the CND (Herrnstein \& Ho 2005).
Our data also could not confirm the connection between them.
\item[Bend of the +20 km s$^{-1}$ cloud]
Zylka et al. (1999) noticed that the northern part of the +20 km s$^{-1}$ cloud bends toward negative velocities.
They also mentioned that the bend seems to combine with the CND.
Our data distinctly delineate this bend and its connection to the NLE with a fine spatial resolution.
\end{description}

\subsection{Line Intensity Ratios}

\begin{figure*}[bht]
\begin{center}
\includegraphics[scale = 1.5]{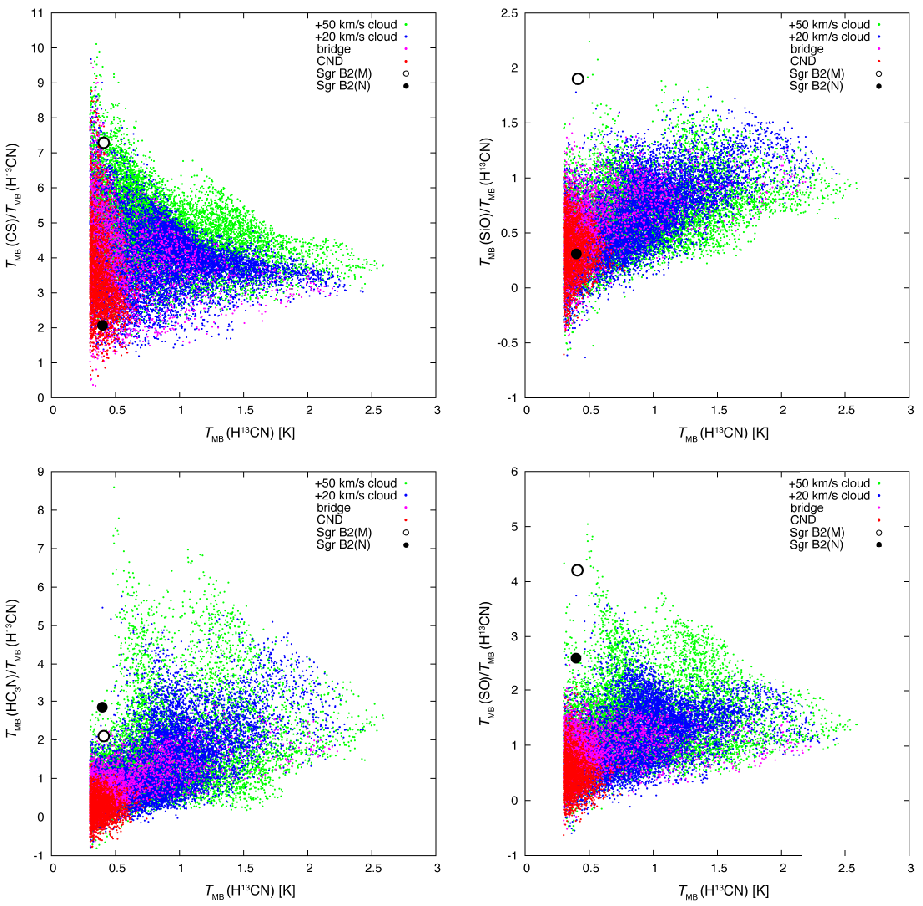}
\caption{Scatter plots of the H$^{13}$CN {\it J}=1--0 intensity ($T_{\rm MB}$) vs. the CS/H$^{13}$CN, SiO/H$^{13}$CN, SO/H$^{13}$CN, and HC$_3$N/H$^{13}$CN ratios.
Green is for the +50 km s$^{-1}$ cloud, blue is for the +20 km s$^{-1}$ cloud, magenta is for the bridge, and red is for the CND.
We defined segments of the +20 and +50 km s$^{-1}$ clouds, the bridge, and the CND as the {\it l}--{\it b}--{\it V} boxes listed in Table 1.
The data were clipped by $3\sigma$ detection of the H$^{13}$CN line (0.3 K).
The black and white filled circles show the ratios toward Sgr B2(N) ($\alpha_{\rm J2000}$=17$^{\rm h}$47$^{\rm m}$20$^{\rm s}$.0, $\delta_{\rm J2000}$=$-$28$\arcdeg$22$\arcmin$19.0$\arcsec$) and Sgr B2(M) ($\alpha_{\rm J2000}$=17$^{\rm h}$47$^{\rm m}$20$^{\rm s}$.4, $\delta_{\rm J2000}$=$-$28$\arcdeg$23$\arcmin$07.0$\arcsec$), respectively.
We used the averaged intensities of the LSR velocity range from +30 to +45 km s$^{-1}$ and from +40 to +50 km s$^{-1}$ for Sgr B2(N) and Sgr B2(M), respectively.
These data were obtained with the IRAM 30 m radio telescope by Belloche et al. (2013).
}
\end{center}
\end{figure*}

Features connected with the CND have been discussed in several works as described above.
Our high-resolution data have manifested the emission continuity between the +20 km s$^{-1}$ cloud and the CND as the bridge, clump, and the NLE for the first time.
The continuity may indicate a physical connection between the +20 km s$^{-1}$ cloud and the NLE.
In order to seek more support for the connection, we examine line intensity ratios.
The {\it J}=1--0 lines of HCN and HCO$^+$ are well-established probes of dense molecular gas.
However, we do not use these lines in the following analysis because they tend to be optically thick for the molecular clouds in the Galactic center and suffer from strong absorption by the foreground.
We use the line intensity ratios of the H$^{13}$CN to the CS, SiO, SO, and HC$_3$N lines to diagnose the chemical difference between the +50 km s$^{-1}$ cloud, the +20 km s$^{-1}$ cloud, the CND, and the bridge.

The CS line is often used as a good probe of dense molecular gas as well as the lines of HCN and HCO$^+$, and CS is not vulnerable to UV photons (Mart\'{i}n et al. 2012).
On the other hand, HC$_3$N can be easily destroyed by UV photons (Rodr\'{i}guez-Franco et al. 1998) and reaction with C$^+$ ions (Meier \& Turner 2005).
Thus the HC$_3$N line is a tracer of dense shielded gas.
Violent shocks can produce SiO through the sputtering of Si-bearing material in grains (Schilke et al. 1997).
The SiO line is a well-established shocked gas probe (e.g., Mart\'in-Pintado et al. 1992).
The abundance of sulfur-bearing molecules, such as CS and SO, may also be enhanced by strong shocks (Harada et al. 2015).

We defined four regions as the {\it l}--{\it b}--{\it V} boxes listed in Table 1 to cover the +20 and +50 km s$^{-1}$ clouds, the bridge, and the CND.
Figure 6 shows plots of the H$^{13}$CN line intensity vs. the CS/H$^{13}$CN, SiO/H$^{13}$CN, SO/H$^{13}$CN, and HC$_3$N/H$^{13}$CN ratios in each region.
The data were clipped by $3\sigma$ detection of the H$^{13}$CN line (0.3 K).
Assuming an abundance ratio of [H$^{12}$CN]/[H$^{13}$CN]=[$^{12}$C]/[$^{13}$C]=24 (Langer \& Penzias 1990), the HCN/H$^{13}$CN ratio suggested that the optical depths of the H$^{13}$CN line are less than 0.2 in the CND and range from 0.1 to 0.8 (mostly less than 0.5) in the GMCs.
Thus, the H$^{13}$CN intensity was used as an indicator of the column density.

In all four panels, the points of the +20 km s$^{-1}$ cloud (blue) are widely scattered, and those of the +50 km s$^{-1}$ cloud (green) are more widely scattered.
There is a common trend that the points of the CND (red) are concentrated on the lower left.
We also list the average ratios (Table 2).
The chemical properties of the GMCs and the CND are distinctly different (Takekawa et al. 2014).
Note that the CS/H$^{13}$CN ratio of the +50 km s$^{-1}$ cloud is typically larger than that of the +20 km s$^{-1}$ cloud.
All of the ratios tend to be significantly smaller in the CND than those in the +20 km s$^{-1}$ cloud.

\begin{deluxetable}{cccc}
\tablecolumns{4}
\tablewidth{0 pc}
\tablecaption{Definitions of the {\it l}-{\it b}-{\it V} boxes for line intensity ratios}
\tablehead{Region & {\it l}  [deg] & {\it b}  [deg] & $V_{\rm LSR}$  [km s$^{-1}$]}
\startdata
50 MC & $-0.06$ to $-0.01$ &  $-0.11$ to $-0.06$ & +25 to +65 \\
20 MC\tablenotemark{a} & $-0.11$ to $-0.06$ &  $-0.11$ to $-0.06$ & +10 to +50 \\
Bridge  & $-0.11$ to $-0.07$ &  $-0.07$ to $-0.03$ & $-30$ to +10 \\
CND & $-0.09$ to $-0.06$ &  $-0.06$ to $-0.03$ & $-130$ to $-30$ \\
  &$-0.06$ to $-0.03$ & $-0.06$ to $-0.03$ & +65 to +130
\enddata
\tablenotetext{a}{
The notations of ``50 MC'' and ``20 MC'' indicate the +50 and +20 km s$^{-1}$ clouds, respectively.}
\end{deluxetable}

For comparison, we also plotted the ratios of Sgr B2(N) and Sgr B2(M), which were calculated based on the results of the line surveys with the IRAM 30 m radio telescope by Belloche et al. (2013)\footnote{We used the public data at \url{http://cdsarc.u-strasbg.fr/viz-bin/qcat?J/A+A/559/A47}.}.
These regions are well-known high-mass star-forming regions (e.g., Gaume \& Claussen 1990).
Unlike plots of Sgr B2, those of the bridge (magenta in Figure 6) are mainly distributed between those of the CND and the +20 km s$^{-1}$ cloud.
The representative ratios are also between them.
These imply that the chemical and physical properties of the bridge may not be extreme but a hybrid of the +20 km s$^{-1}$ cloud and the CND.
The intermediate line ratios may be attributed to the possible physical connection between the +20 km s$^{-1}$ cloud and the CND.

\begin{deluxetable}{ccccc}
\tablecolumns{5}
\tablewidth{0 pc}
\tablecaption{Ratios between integrated line intensities}
\tablehead{Ratio & 50 MC & 20 MC\tablenotemark{a}& Bridge & CND}
\startdata
 CS/H$^{13}$CN & 4.59 & 3.90 & 3.83 & 3.36 \\
 SiO/H$^{13}$CN & 0.69 & 0.75 & 0.64 & 0.33 \\
 SO/H$^{13}$CN & 1.40 & 1.24 & 0.95 & 0.49 \\
 HC$_3$N/H$^{13}$CN &  1.71 & 1.70 & 1.02 & 0.27
\enddata
\tablenotetext{a}{
The notations of ``50 MC'' and ``20 MC'' indicate the +50 and +20 km s$^{-1}$ clouds, respectively.}
\end{deluxetable}

\subsection{Cloud-Plunging Scenario}

\begin{figure*}[bth]
\begin{center}
\includegraphics[width=17cm]{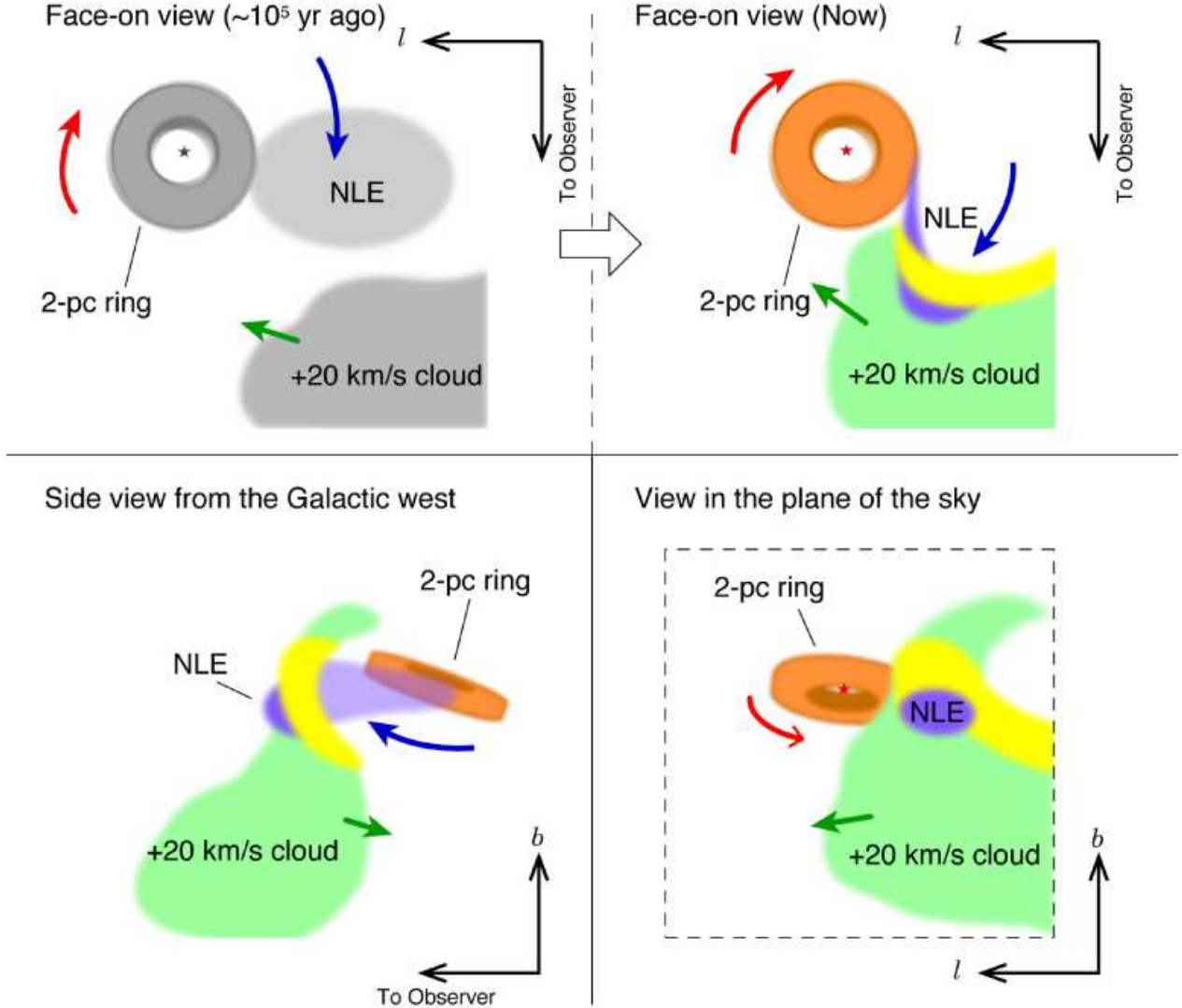}
\vspace{5mm}
\caption{Three-dimensional schematic view of the cloud-plunging scenario.
The 2 pc ring is depicted in orange, the NLE in purple, +20 km s$^{-1}$ cloud in green, and the bridge in yellow.  
The small star indicates the location of Sgr A$^*$.
The upper-left panel in shows the face-on view $\sim 10^5$ yr ago and the upper-right panel shows the current face-on view.  
The black orthogonal arrows indicate the directions toward the observer and the positive Galactic longitude.  
The red and blue arrows represent rotational motion of the CND.
The lower-left panel shows the current side view and the lower-right panel shows the current sky view. 
The green arrows represent the possible bulk motion of the +20 km s$^{-1}$ cloud and the dashed-line square is our observation area. }
\end{center}
\end{figure*}

The emission continuity of the +20 km s$^{-1}$ cloud and the CND may indicate physical contact between the +20 km s$^{-1}$ cloud, the bridge, and the CND. 
Because it is known that the +20 km s$^{-1}$ cloud is in front of the Galactic nucleus (e.g., G\"{u}sten \&\ Henkel 1983), the bridge may be thought of as a streamer from the +20 km s$^{-1}$ cloud toward the nucleus.  
{ However, the upper part of the +20 km s$^{-1}$ cloud seems to be accelerating toward us.
 This fact} is incompatible with the streamer interpretation of the bridge.
This invokes the cloud-plunging scenario, in which a cloud has plunged into the +20 km s$^{-1}$ cloud.
Figure 7 shows the schematic views of the scenario.
The NLE, which is the asymmetric part of the CND, may be a remnant of the parent GMC that has been captured and disrupted by the potential well at the nucleus (Mapelli \& Trani 2016).
This asymmetric part may have plunged into the northern part of the +20 km s$^{-1}$ cloud $\sim 10^5$ years ago.
It compressed, accelerated, and swept up molecular gas in the +20 km s$^{-1}$ cloud that had been just in front of the NLE, to form the connecting bridge and the arc of emission.
The negative-velocity end of the NLE, which may correspond to its ``head", has an LSR velocity similar to the rotational velocity of the 2 pc ring.
Thus, the head of the NLE may have already gone through the northern periphery of the +20 km s$^{-1}$ cloud.
Some fractions of the progenitor of the NLE with smaller angular momentum may have gone ahead on inner orbits without colliding with the +20 km s$^{-1}$ cloud.
This streaming gas may compose the Western Streamer (Liu et al. 2012).

\begin{figure*}[!tbh]
\begin{center}
\includegraphics[scale=2.0]{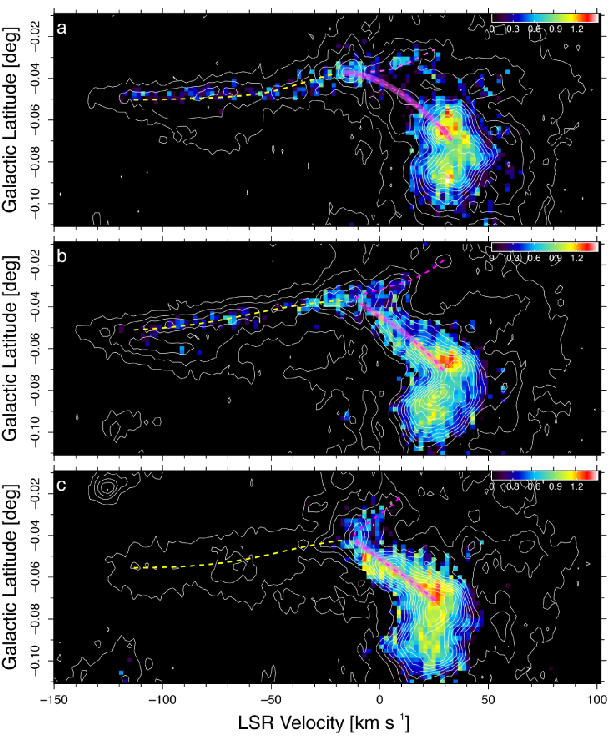}
\caption{
Distribution of the SiO {\it J}=2--1/H$^{13}$CN {\it J}=1--0 intensity ratio in the same {\it b--V} planes as shown in Figure 4 (a, b, c).
The yellow and magenta lines are the same as in Figure 4. The white contours show the CS {\it J}=2--1 intensities at 0.5 K intervals from the 0.3 K minimum. 
}
\end{center}
\end{figure*}

Because the difference in velocity between the NLE head and the +20 km s$^{-1}$ cloud exceeds 100 km s$^{-1}$, it is expected with the cloud-plunging scenario that the gas-phase abundance of refractory molecules will increase in the shocked layer of the colliding clouds (e.g., Habe \& Ohta 1992).
SiO is a well-established shocked gas probe (e.g., Mart\'in-Pintado et al. 1992; Schilke et al. 1997).
Figure 8 shows the {\it b--V} maps of  the SiO {\it J}=2--1/H$^{13}$CN {\it J}=1--0 intensity ratio at the same Galactic longitudes as in Figure 4.
{ The ratio was calculated for the locations where detection of the H$^{13}$CN line was $\geq 3\sigma$ (0.3 K).
We found that} the ratio is enhanced ($\gtrsim 1$) in the northern edge of the +20 km s$^{-1}$ cloud, where the Southern Streamer lies (Okumura et al. 1989, 1991; Coil \& Ho 1999, 2000).
Liu et al. (2013) reported that the SiO {\it J}=1--0/C$^{34}$S {\it J}=1--0 intensity ratio was also significantly enhanced in the northern edge of the +20 km s$^{-1}$ cloud (see Figure 9 in their paper).
These enhancements may be due to the increase in SiO abundance caused by shocks.
The authors suggested the presence of infalling clouds on noncircular orbits (Liu et al. 2012).
Wright et al. (2001) also suggested that the Southern Streamer also experienced shock heating, which was possibly induced by the supernova of Sgr A East.
Although any of these shock processes including ours cannot currently be ruled out, we interpret that the SiO enhancement at the root of the bridge may be caused by the putative violent collision of the clouds, thereby providing support for the cloud-plunging scenario.

\subsection{Implication for Gas Feeding}
The collision between the +20 km s$^{-1}$ cloud and the NLE cancels out their angular momentum and they lose kinetic energy with dissipation of the shock.
As a result, part of the NLE may accrete inward along highly eccentric orbits.
These processes may lead to growth of the 2 pc ring and possibly feed the inner cavity.

We estimated that the gas mass from this collision available for feeding is $\sim10^4\, M_{\odot}$ using the CS {\it J}=2--1 data in the box defined by $-0.09\arcdeg\!\leq\!l\!\leq\!-0.07\arcdeg$, $-0.06\arcdeg\!\leq\!b\!\leq\!-0.03\arcdeg$, and $-80\!\leq\!V_{\rm LSR}\!\leq\!0$ km s$^{-1}$.
For the estimation, we assumed local thermodynamic equilibrium (LTE) with an excitation temperature of 10 K (Tsuboi et al. 2015), an optical depth of 0.4, a $[\rm CS]/[H_2]$ fractional abundance of $10^{-8}$ (Requena-Torres et al. 2006), and a distance to the Galactic center of 8.3 kpc (Gillessen et al. 2009).
The optical depth was estimated from the CS {\it J}=2--1/$^{13}$CS {\it J}=2--1 intensity ratio obtained by our previous line surveys (Takekawa et al. 2014), assuming an abundance ratio of [$^{12}$CS]/[$^{13}$CS]=[$^{12}$C]/[$^{13}$C]=24 (Langer \& Penzias 1990).
The mass of the CND was estimated from the CS intensity to be $\sim 10^5\, M_{\odot}$.
This is comparable to the mass of (2--5)$\times$10$^5$ $M_{\odot}$ that was estimated by Oka et al. (2011).
The available mass ($\sim 10^4\, M_{\odot}$) is one order of magnitude smaller than the current mass of the CND.
Therefore, the possible feeding of the CND does not significantly alter its Toomre's $Q$ parameter ($Q\equiv\kappa\sigma_{\rm V}/\pi G\Sigma_{\rm gas}$; Toomre 1964) of $\sim$30, so the CND remains sufficiently stable against self-gravitational fragmentation. 
Nevertheless, it is still possible that this collision increases the mass-feeding rate to the central 1 pc if the infalling gas reaches the inner cavity without accreting onto the CND.

Our cloud-plunging scenario naturally explains the observed kinematics and proposes a possible pathway for the transportation of the molecular gas from the outer GMCs to  within the central few parsecs.
This scenario could be bolstered with direct measurements of the three-dimensional kinematics of the region, such as that performed for the minispiral with the Very Large Array (VLA; Zhao et al. 2009).
The long-baseline observations performed with the Atacama Large Millimeter/submillimeter Array (ALMA) at $\sim20$ year intervals will be able to measure the proper motions of molecular clumps and thereby delineate the three-dimensional kinematics.

\section{Summary}
This paper presents the results of mapping observations of the central 10 pc of our Galaxy in seven molecular lines at millimeter wavelength with the NRO 45 m telescope.
The principal results of this study are summarized as follows:
\begin{enumerate}
\setlength{\leftskip}{4mm}
\item An emission bridge that may connect the +20 km s$^{-1}$ cloud (M--0.13--0.08) and the NLE of the circumnuclear disk (CND) was detected in the CS, H$^{13}$CN, SiO, SO, and HC$_3$N lines.

\item The bridge has intermediate line intensity ratios of CS/H$^{13}$CN, SiO/H$^{13}$CN, SO/H$^{13}$CN, and HC$_3$N/H$^{13}$CN between those of the +20 km s$^{-1}$ cloud and the CND.
This is consistent with the interpretation that the chemical and physical properties of the bridge are a hybrid of the +20 km s$^{-1}$ cloud and the CND.

\item The cloud-plunging scenario, in which the NLE plunged into the +20 km s$^{-1}$ cloud, was proposed to explain the kinematics of the bridge, the NLE, and the bend of the +20 km s$^{-1}$ cloud.

\item Such collisional events can feed the CND and thereby could promote mass feeding into the central parsec.

\end{enumerate}

\acknowledgments
We are grateful to the Nobeyama Radio Observatory (NRO) staff for their excellent support of the 45 m telescope observations.
The NRO is a branch of the National Astronomical Observatory of Japan, National Institutes of Natural Sciences.
We thank the anonymous referee for helpful comments and suggestions that improved this paper.
This study was supported by a Grant-in-Aid for Research Fellow from the Japan Society for the Promotion of Science (15J04405).

\end{document}